\documentclass{ws-procs975x65}

\begin{document}

\title{MG13 proceedings: On the validity of the adiabatic
  approximation in compact binary inspirals}

\author{Andrea Maselli, Leonardo Gualtieri, Valeria Ferrari}
\address{Dipartimento di Fisica, "Sapienza", Universit\`a di Roma \&
  Sezione INFN Roma1, Rome, Italy.}

\author{Francesco Pannarale}

\address{Max-Planck-Institut f{\"u}r Gravitationsphysik, Albert
  Einstein Institut, Potsdam, Germany.}

\begin{abstract}
  We use the post-Newtonian-Affine model to assess the validity of the
  adiabatic approximation in modeling tidal effects in the phase
   evolution of compact binary systems. We compute the dynamical
  evolution of the tidal tensor, which we estimate at the $2\,$PN
  order, and of the quadrupole tensor, finding that their ratio,
  i.e. the tidal deformability, increases in the last phases of the
  inspiral. We derive the gravitational wave phase corrections
  due to this phenomenon and quantify how they affect
  gravitational wave detectability.
\end{abstract}


\bodymatter

\section{Introduction}
\label{sec:intro}


Coalescing binary systems composed of neutron stars (NS) and/or black
holes (BH) are one of the most promising sources of gravitational
waves (GWs) for second and third generation of ground based
interferometers \cite{VL}. Additionally, certain key features of
NS tidal deformations may provide precious information about the NS
equation of state (EOS)\cite{FH08,HLLR10,DN09,PROR11}. In current
semi-analytical approaches, finite size effects are treated by
assuming that the NS deformability is described in terms of a set of
quantities, the {\it Love numbers} \cite{H08,BP09}, among which
  the apsidal constant $k_{2}$ is the most important one for GW
  phenomenology. Within this framework and under the so called
  {\it adiabatic approximation}, the evolution of a star embedded
  in an external quadrupolar tidal field, $C_{ij}$, is governed by
the equation
\begin{equation}
\label{ad}
Q_{ij}=-(2/3)k_2R_\text{NS}^5C_{ij}\,,
\end{equation}
where $Q_{ij}$ and $R_\text{NS}$ are the star quadrupole moment
  and radius at isolation, respectively.


In this work, on the basis of the results of Maselli et
al.~2012\cite{MGPF12}, we employ the post-Newtonian-Affine model
(PNA)\cite{FGM11}, recently developed to study tidal interactions in
BH-NS and NS-NS systems, to asses the range of validity of
Eq.\,(\ref{ad}). By computing the ratio $\sim Q_{ij}/C_{ij}$ as
function of the orbital separation for several binary
configurations, we find that $k_{2}$ is not constant during the
inspiral and that it grows in the last stages of the inspiral. 
Finally, we estimate the impact of such effect on future GWs detections.


\section{The PNA model}
\label{sec:PNA_model}


To describe stellar deformations, we improve the PNA model, based on
the Affine approach \cite{CL85,LM85,WL00,FGP09, FGP10}, which relies
on the assumption that a NS in a binary system is deformed into an
ellipsoid by the tidal forces exerted by its companion. Under the
Affine hypothesis, the infinite degrees of freedom of the stellar
fluid reduce to five dynamical variables, namely the three
principal axes of the ellipsoid and two angles related to the star
angular velocity and the internal fluid motion. The main improvements
to the original Affine approach \cite{CL85,LM85,WL00}, were
brought in \cite{FGP09,FGP10,FGM11}, where the NS spherical
equilibrium configuration was determined by solving the
relativistic stellar structure equations, and a more careful
treatment of coordinate systems and of relativistic corrections to the
star internal dynamics were introduced. Furthermore,
starting from a two-body post-Newtonian metric\cite{BFP98,FBA06}, an
explicit expression for the tidal field was derived. This
represents a major change from the original approach in which the NS
moves in a background space-time described by the Kerr metric,
and it allows to consistently treat stellar deformations in NS-NS
binaries.
The orbital motion is handled within a post-Newtonian framework,
as well: the phase evolution is determined by means of the TaylorT4
approximant\cite{SOA10}, including the effects of tidal
interactions on the orbital motion, up to $1\,$PN
order\cite{VFH11}.


The source of stellar deformations in the PNA model is given by
the quadrupolar tidal tensor
$C_{ij}=R_{\alpha\beta\gamma\delta}e^{\alpha}_{(0)}e^{\beta}_{(i)}e^{\gamma}_{(0)}e^{\delta}_{(j)}$,
where $R_{\alpha\beta\gamma\delta}$ is the Riemann tensor of the
$3\,$PN metric describing the orbital motion, and $e_{(i=0,\ldots,3)}$
is an orthonormal tetrad field fixed to the star center of mass
and parallel transported along its motion. The results obtained
in \cite{FGM11}, where we derived $C_{ij}$ up to
$\mathcal{O}(1/c^3)$, were extended in \cite{MGPF12} to include
$\mathcal{O}(1/c^4)$ terms. With this aim, we also
computed the $2\,$PN contribution to the spatial tetrad
vectors $e^{j}_{(k)}$ introduced in \cite{F88}. $C_{ij}$ depends on
the masses of both compact objects $m_{1,2}$, and includes,
$1.5\,$PN order, the spin contribution from rotating objects.


\section{Tidal Love number evolution}
\label{sec:GW_wave}

We solve the PNA equations of motion up to the onset of the NS mass
shedding, for a representative set of binary system configurations, by
considering two equations of state which are expected to cover the
range of possible NS deformabilities. For each binary
  model, we compute $k_2$ as the ratio between the quadrupole and the
tidal tensor, from Eq.\,(\ref{ad}). We find that this quantity is
a function of the orbital separation $r$, i.e.~$k_2=k_2(r)$, that
increases during the binary inspiral by a $\sim10\%-30\%$
factor. The constant Love number $\bar{k}_2$ should be regarded as
 the asymptotic limit of this function, i.e.  $\bar k_2=\lim_{r\rightarrow\infty} k_2(r)\,.$ 
This effect is strongly dependent on the EOS choice: it is larger/smaller for a 
less/more compact NS. The dependence of this effect on the mass ratio is weaker,
and it increases more for larger values of $q=m_{2}/m_{1}\leq 1$.
This effect appears to be nearly insensitive to the NS mass.


Replacing the Love function $k_{2}(r)$ into the expression of the GW
signal, we estimate a new correction to the GW phase. This adds 
linearly to the point-particle (PP) term, and to the tidal
term (T) computed by assuming the Love number to be constant:
\begin{equation}
\psi_{\textnormal{GW}}=\psi_{\textnormal{PP}}+\psi_{\textnormal{T}}+\delta\psi_{T}\ ,
\end{equation}
where $\delta\psi_{T}$ describes the growth of $k_{2}(r)$ during the
inspiral. We find that the new contribution is always less than $1$ radian for all
considered configurations, up to the onset of mass shedding.


To assess the validity of the adiabatic approximation as far GW
detectability is concerned, we compute overlaps \cite {LOB08} between
PP templates and {\it real signals} which take into account tidal
effects at the best of our knowledge, i.e.~including
$\delta\psi_{T}$. We find that in the worst case, i.e.~for a NS-NS
binary with a standard total mass of $2.8M_{\odot}$, the number
of missed events due to the use of PP templates could be as high
as $11$\% and $13$\% for the second and third generation of ground
based detectors, respectively, depending on the NS equation of
state. By using templates which include tidal corrections by
means of the constant Love number, however, the overlaps would
differ from unity by less than a part in one thousand for all
configurations considered. We may thus conclude that
building gravitational waveforms within the adiabatic approximation is
very reliable up to the mass-shedding.

\section{Conclusions}
\label{sec:disc}


In this work we reviewed the results of \cite{MGPF12}, in
  which we improved the PNA model by computing the $2\,$PN
contribution to the post-Newtonian tidal tensor. We used this
framework to assess the reliability of the Love number adiabatic
approximation and found that the NS deformability increases
during the last phases of the inspiral up to $30$\%, depending on the
NS EOS.



\end{document}